\documentclass[acmlarge]{acmart}

\AtBeginDocument{%
  }

\setcopyright{acmlicensed}
\copyrightyear{2024}
\acmYear{2024}
\acmDOI{XXXXXXX.XXXXXXX}

\acmISBN{978-1-4503-XXXX-X/18/06}

\usepackage{colortbl} 
\usepackage{geometry} 
\begin{document}

\title{Dr. GPT in Campus Counseling: Understanding Higher Education Students’ Opinions on LLM-assisted Mental Health Services}


\author{Owen Xingjian Zhang}
\affiliation{%
  \institution{Princeton University}
  \streetaddress{1 Th{\o}rv{\"a}ld Circle}
  \city{Princeton}
  \country{USA}}
\email{owenz@princeton.edu}

\author{Shuyao Zhou}
\affiliation{%
  \institution{Princeton University}
  \streetaddress{1 Th{\o}rv{\"a}ld Circle}
  \city{Princeton}
  \country{USA}}  
\email{sz8740@princeton.edu}

\author{Jiayi Geng}
\affiliation{%
  \institution{Princeton University}
  \streetaddress{1 Th{\o}rv{\"a}ld Circle}
  \city{Princeton}
  \country{USA}}
\email{jiayig@princeton.edu}

\author{Yuhan Liu}
\affiliation{%
  \institution{Princeton University}
  \streetaddress{1 Th{\o}rv{\"a}ld Circle}
  \city{Princeton}
  \country{USA}}
\email{yuhanl@princeton.edu}

\author{Sunny Xun Liu}
\affiliation{%
  \institution{Stanford University}
  \streetaddress{1 Th{\o}rv{\"a}ld Circle}
  \city{Stanford}
  \country{USA}}
\email{sunnyxliu@stanford.edu}

\renewcommand{\shortauthors}{Zhang et al.}




\begin{abstract}
In response to the increasing mental health challenges faced by college students, we sought to understand their perspectives on how AI applications, particularly Large Language Models (LLMs), can be leveraged to enhance their mental well-being. Through pilot interviews with ten diverse students, we explored their opinions on the use of LLMs across five fictional scenarios: General Information Inquiry, Initial Screening, Reshaping Patient-Expert Dynamics, Long-term Care, and Follow-up Care. Our findings revealed that students' acceptance of LLMs varied by scenario, with participants highlighting both potential benefits, such as proactive engagement and personalized follow-up care, and concerns, including limitations in training data and emotional support. These insights inform how AI technology should be designed and implemented to effectively support and enhance students' mental well-being, particularly in scenarios where LLMs can complement traditional methods, while maintaining empathy and respecting individual preferences.
\end{abstract}

\maketitle

\section{Introduction}
Today’s young adults, including higher education students, are reporting increasingly high levels of depressive symptoms, stress, and loneliness, surpassing those of older cohorts \cite{beiter2015prevalence, blanco2008mental}. Studies link these mental health issues to academic pressures, future career concerns, achievement culture, the COVID-19 pandemic, and a lack of mental health resources \cite{duffy2019trends, eisenberg2007help, wang2020investigating}. On average, it takes about 7.8 days for college students to get an initial appointment with a mental health professional, but the following sessions could extend to several weeks due to a shortage of mental health services on campuses \cite{healthyminds2023}. This shortage is reflected in the counselor-to-student ratio, which often falls short of the recommended 1:500 standard, with many colleges having only one counselor for every 1,000 to 1,500 students \cite{cornett2023}. In addition to the shortage of mental health resources, college students are reluctant to seek traditional treatment for multiple reasons, such as financial cost, time constraints, and concerns about stigma \cite{hunt2010mental}, . These findings highlight the urgent need for innovative solutions, such as technology, to address mental health challenges in this demographic. Researchers are considering LLM-powered chatbots \cite{lai2023psyllm} for mental health support. Large Language Models (LLMs) \cite{brown2020language} are advanced AI systems capable of understanding and generating human-like text, which can provide personalized mental health support by engaging in therapeutic conversations, diagnosing conditions, performing semantic analysis, and searching therapy data, thereby enhancing mental health interventions \cite{ricks2023therapy, smith2023semantic}.

College students can be an ideal demographic population to explore the impact of LLMs for mental health services as they are increasingly using LLM applications for academic purposes \cite{perkins2023academic} and showing a growing demand for mental health services \cite{gulliver2012barriers}. However, to our knowledge, there is no empirical study that has investigated higher education students’ perceived potential benefits and concerns regarding using LLMs in the mental health domain. Based on related work, to study students’ opinions on potential LLM use in mental health services, we designed five fictional scenarios in General Information Inquiry, Initial Screening, Reshaping Patient-Expert Dynamics, Long-term Care, and Follow-up Care.

\section{Related Work}
LLMs present significant opportunities and challenges in the field of mental health services \cite{ma2023understanding}. They offer the potential to enhance accessibility, improve efficiency, and maintain privacy through automated interactions. Previous research suggests that LLMs can support mental health in five distinct approaches: General Information Inquiry, Initial Screening, Reshaping Patient-Expert Dynamics, Long-term Care, and Follow-up Care. However, concerns about their reliability, the potential to generate unpredictable responses, and their ability to provide emotional support as effectively as human interactions remain significant challenges \cite{jo2023understanding, montemayor2022principle}.

\textbf{General Information Inquiry:} Research has shown that computational or predictive analyses of digital data can accurately identify moods and mental health states \cite{de2013predicting, de2013predictingDepression}. These capabilities lay the foundation for LLMs to provide informative support, crucial in this scenario. However, there is a risk of LLMs propagating misinformation due to reliance on online data for fact-checking, which could lead to incorrect assessments or advice \cite{bender2021dangers, bhatt2021universality}.

\textbf{Initial Screening:} The shortage of mental health professionals and the increasing dependence on technology for mental health services highlight the importance of LLMs in offering initial screening and basic counseling \cite{moore2019bot, boucher2021artificially}. Yet, concerns about the depth of emotional support that LLMs can provide persist, questioning their effectiveness in replacing human interactions \cite{montemayor2022principle}.

\textbf{Reshaping Patient-Expert Dynamics:} The integration of chatbots in mental health services promises to extend accessibility and reshape traditional healthcare dynamics \cite{eysenbach2023role}. Nonetheless, the potential for generating unpredictable responses could undermine the trust required in patient-expert relationships \cite{bender2021dangers}.

\textbf{Long-term Care:} The ability of LLMs to adjust and evolve in response to user interactions is essential for their effectiveness and long-term sustainability in therapeutic settings \cite{thorp2023chatgpt}. However, the reliance on potentially biased online data poses a challenge, as it may lead to the reinforcement of incorrect or harmful patterns \cite{bender2021dangers}.

\textbf{Follow-up Care:} Previous studies have shown that chatbots can effectively deliver follow-up care and maintain continuity through coherent responses to user prompts \cite{schleider2022randomized}. Still, the limitations in providing nuanced emotional support could impact the quality of ongoing care \cite{montemayor2022principle}.

\section{Pilot Results}
We conducted a pilot study using semi-structured interviews with participants to evaluate their acceptance of LLMs across five different scenarios in mental health services. This study revealed varied levels of acceptance for LLMs across these scenarios, providing critical insights into where these technologies may be most effective in supporting mental health services. Participants responded most positively to the "Initial Screening" and "Follow-up Care" scenarios, which were seen as areas where LLMs could significantly enhance the accessibility and efficiency of care.

\begin{table*}[ht]
\centering
\caption{Summary of Scenarios: Acceptance Levels, Potential Benefits, and Concerns}
\label{summary-table}
\begin{tabular}{|>{\columncolor{gray!20}}c|>{\columncolor{gray!20}}m{2.3cm}|>{\columncolor{gray!20}}m{1.5cm}|>{\columncolor{gray!20}}m{5.2cm}|>{\columncolor{gray!20}}m{5.2cm}|}
\hline
\textbf{\#} & \textbf{Scenario} & \textbf{Acceptance Level} & \textbf{Potential Benefits} & \textbf{Concerns} \\ \hline
1 & General Informative Support & Medium & Proactive engagement with patients, identifying mental health symptoms, emotion detection & Risk of incorrect or unempathetic responses, lack of specific training for mental health data \\ \hline
2 & Initial Screening & High & Customized questions based on professional questionnaires, helping students express their mental health states & Handling complex emotional states, reliability of data \\ \hline
3 & Reshaping Patient-Expert Dynamics & Low & Preferences for LLMs as background tools, nurses, virtual assistants, visualizers & Lack of emotional support, discomfort, unwillingness to share personal information, high risks in counseling sessions \\ \hline
4 & Long-term Care & Medium & Continuously improving responses based on user interactions, enhancing non-verbal cue detection & Quality of long-term interactions, balancing usefulness and bias, handling complex human judgments \\ \hline
5 & Follow-up Care & High & Convenience and efficiency in follow-up care, embedding visual aids to enhance understanding, understanding richer information from non-verbal cues & Privacy and accuracy of follow-up information, lack of responsibility for incorrect outputs \\ \hline
\end{tabular}
\end{table*}

\textbf{Initial Screening:} In the Initial Screening scenario, LLMs were highly accepted by participants for their ability to personalize interactions and tailor questions to the specific needs of users. This customization was identified as a key strength, allowing students to express their mental health concerns more effectively and facilitating a more engaging and responsive screening experience. Participants appreciated how LLMs could streamline the initial stages of mental health support by providing timely and relevant responses, potentially expediting access to further care. The positive reception of LLMs in this scenario underscores their potential to enhance the efficiency and effectiveness of mental health screenings, making them a valuable tool in early intervention efforts.

\textbf{Follow-up Care:} The Follow-up Care scenario also received high acceptance, with participants recognizing the value of LLMs in maintaining continuity of care through regular check-ins and reminders. This consistent engagement was viewed as particularly beneficial in supporting adherence to treatment plans and providing ongoing motivation for users. Participants noted that the ability of LLMs to offer timely follow-up support could help bridge gaps in traditional care pathways, ensuring that individuals continue to receive the attention they need between sessions with a human provider. As with initial screening, it is essential that LLMs in follow-up care are designed to complement human involvement, with the capacity to escalate concerns to professionals when necessary. This approach ensures that the emotional and relational aspects of care are preserved, even as technology is leveraged to enhance service delivery.

\section{Discussion}
The findings from our pilot study across all five scenarios provide essential insights into how LLMs can be effectively integrated into mental health services to enhance students' mental well-being. These insights inform the strategic design and implementation of AI technology, ensuring that it complements traditional methods while maintaining empathy and respecting user preferences.

\textbf{Designing AI to Complement Human Care:} A consistent theme across the scenarios was the importance of designing LLMs to enhance, rather than replace, human care. In scenarios like Initial Screening and Follow-up Care, where LLMs were highly accepted, their ability to improve accessibility and efficiency was clear. Participants appreciated how LLMs could manage routine tasks, offer personalized support, and provide timely assistance, particularly when human resources are limited. However, the success of these technologies hinges on their ability to work in tandem with human providers, recognizing the limits of AI in handling complex emotional issues \cite{boucher2021artificially, montemayor2022principle}. LLMs must be equipped with mechanisms to escalate cases to human professionals when necessary, ensuring that their use enhances service quality without compromising the therapeutic relationship. Moreover, in scenarios like General Information Inquiry and Long-term Care, the accuracy and reliability of AI-generated content are crucial. By maintaining high standards of information accuracy and incorporating features that enable seamless collaboration between AI and human experts, LLMs can significantly contribute to more effective mental health support \cite{bender2021dangers}.

\textbf{Maintaining Empathy and User-Centered Design:} Another critical consideration is the need to maintain empathy and personalization in mental health care. Across all scenarios, participants expressed the importance of LLMs being able to provide emotionally intelligent, personalized care, particularly in sensitive contexts like Reshaping Patient-Expert Dynamics and Long-term Care \cite{moore2019bot, montemayor2022principle}. To address these concerns, LLMs must be designed with advanced emotional intelligence, allowing them to respond to the nuanced needs of students effectively. Additionally, transparency about the AI's role in care is essential for building trust. Ensuring that users understand the capabilities and limitations of AI, and keeping human providers at the center of the therapeutic process, is crucial for fostering a positive relationship between students and AI \cite{eysenbach2023role}. By focusing on empathetic, user-centered design, we can create AI tools that not only improve accessibility and efficiency but also enhance the overall quality of mental health care for students. This approach ensures that LLMs serve as supportive, trustworthy tools in the mental health journey, contributing positively to students' mental well-being \cite{jo2023understanding}.

\bibliographystyle{ACM-Reference-Format}
\bibliography{main}

\end{document}